\documentclass{article}
\usepackage{mrs2005,epsfig}
\usepackage{amsmath}
\usepackage{amsfonts}
\usepackage{graphicx}
\usepackage{graphics}
\usepackage{epsfig}
\usepackage{rotate}

\begin{document}
\title{CHAOS AND THE DYNAMICAL EVOLUTION OF BARRED GALAXIES}

\author{T. Manos$^{\dag,\ast}$ and E. Athanassoula$^{\dag}$}{ }
\affil{$^{\dag}$ Observatoire Astronomique de Marseille-Provence
(OAMP), FRANCE.\\
$^{\ast}$ Center for Research and Applications of Nonlinear Systems (CRANS),\\
Department of Mathematics, University of Patras, GREECE.}

\begin{abstract}
The dynamical evolution of barred galaxies depends crucially on the
fraction and their spacial distribution of chaotic orbits in them.
In order to distinguish between the two kinds of orbits, we use the
Smaller Alignment Index (SALI) method, a very powerful method which
can be applied to problems of galactic dynamics. Using model
potentials, and taking into account the full 3D distribution of
matter, we discuss how the distribution of chaotic orbits depends on
the main model parameters, like the mass of the various components
and the bar axial ratio.

\end{abstract}

\section{Introduction}
The distinction between ordered and chaotic motion in dynamical
systems is fundamental in many areas of applied sciences. This
distinction is particularly difficult in systems with many degrees
of freedom (dof), basically because it is not feasible to visualize
their phase space. Thus, we need fast and accurate tools to give us
information about the chaotic or ordered character of the orbits,
especially for conservative systems. In this work we focus our
attention on the method of the \textbf{S}maller \textbf{AL}ignment
\textbf{I}ndex (\textbf{SALI}) \cite{sk:1}, or, as elsewhere called
\textbf{A}lignment \textbf{I}ndex \textbf{(AI)} \cite{VKS} and we
present some applications of the index in Ferrers barred galaxy
potentials of 2 and 3 dof. In order to compute the SALI for a given
orbit, one has to follow the time evolution of the orbit itself and
also of two deviation vectors $v_{1}$ and $v_{2}$, which initially
point in two different directions. At every time step the two
deviation vectors are normalized and the SALI is then computed as:
\begin{equation}\label{eq:1}
    SALI(t)=min\{\|\frac{v_{1}(t)}{\|v_{1}(t)\|}+\frac{v_{2}(t)}{\|v_{2}(t)\|}\|,\|\frac{v_{1}(t)}{\|v_{1}(t)\|}-\frac{v_{2}(t)}{\|v_{2}(t)\|}\|\}.
\end{equation}
In 2 and 3 dof Hamiltonian systems the distinction between ordered
and chaotic motion is easy because the ordered motion occurs on a 2D
or 4D torus, respectively, to which any initial deviation vector
becomes almost tangent after a short transient period. The behavior
of SALI is also discussed in \cite{sk:2}, where some applications
were also presented.

Our goal is to study how the fraction of the ordered and chaotic
trajectories, in 2 and 3 dof, depends on some major parameters of
the models. We use a 2 dof and 3 dof barred potential, which rotates
around its z-axis. The system is rotating with an angular speed
$\Omega_{b}$ and can be described by the following Hamiltonian form:
\begin{equation}\label{eq:2}
    H=\frac{1}{2}(p_{x}^{2}+p_{y}^{2}+p_{z}^{2})+V(x,y,z)-\Omega_{b}(xp_{y}-yp_{x}).
\end{equation}

Our model potential consists of a Miyamoto disc, a Plummer sphere
and a Ferrers bar.

\section{Applications in the 2D case}
We first apply the SALI index to the 2 dof barred potential. In this
case, we can have a Poincar\'{e} Surface of Section (PSS) and can
check the effectiveness of the SALI, by comparing the results.

In figure 1, we present two orbits of different kind: (i)
$(x,y,p_{x},p_{y})=(1.5,0,p_{x}(H),0)$ and (ii)
$(x,y,p_{x},p_{y})=(-0.9,0,p_{x}(H),0)$. For this application
$H=-0.3$. In panels a) and b) of figure 1, we show the orbit
projections in the $(x,y)$-plane and in panel c) we draw their
corresponding PSS, where we can see that the chaotic orbit (i) tends
to fill with scattered points the available part of the plane
$(x,\dot{x})$ and that the ordered orbit (ii) creates a closed
invariant curve. Finally, in panel d) we apply the SALI method for
these two orbits. For the chaotic one, the SALI tends to zero
$(10^{-16})$ exponentially after some time steps while for the
regular orbit, it fluctuates around a positive number. By choosing
initial conditions on the line $\dot{x}=0$ of the PSS and
calculating the values of the SALI, we can detect very small regions
of stability that can not be visualized easily by the PSS method.
Repeating this for many values of the energy, we are able to follow
the change of the fraction of chaotic and ordered orbits in the
phase space as the energy of the system varies.

\section{Applications in the 3D case}
A similar study can be made for the 3 dof case of the barred
potential. We first try a basic model and we vary two parameters,
the length of the short z-axis $c$ and the mass $M_{bar}$ of the
bar. The initial conditions are given in two ways: a) in the plane
$(x,p_{y},z)$ with $(y,p_{x},p_{z})=(0,0,0)$ and b) in the plane
$(x,p_{y},p_{z})$ with $(y,z,p_{z})=(0,0,0)$. In both cases, we find
similar results. As the mass $M_{bar}$ of the bar increases, the
percentage of the chaotic orbits increases as well. This is in good
agreement with the results found for 2 dof by \emph{Athanassoula et
al.} (1983). In the other case, we find that when the length of the
short z-axis $c$ of the bar increases (keeping the $M_{bar}$
constant), the system presents more regular behaviour than in the
initial model.

\begin{figure*}
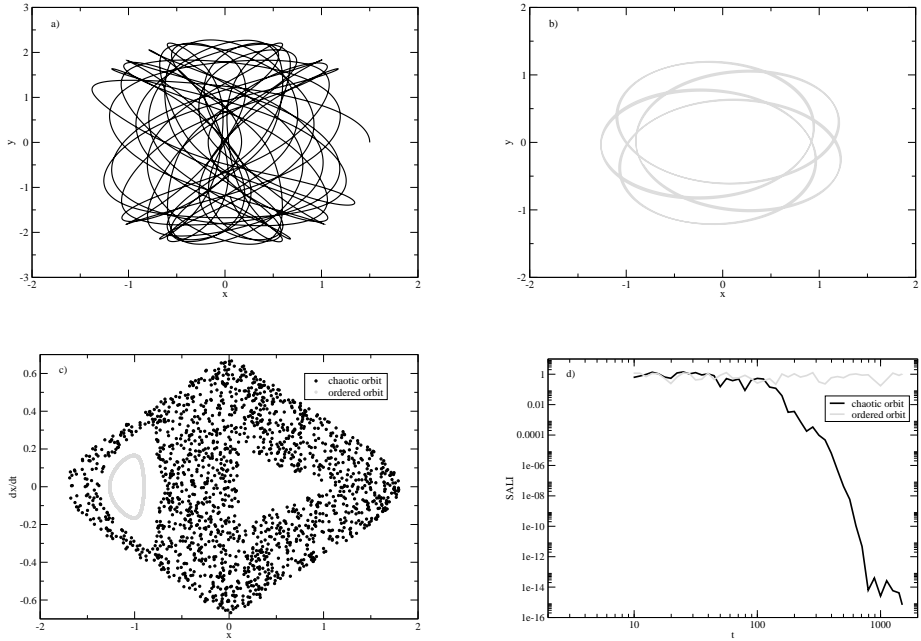

\begin{center}
\epsfig{figure=chao_orb.eps,width=5.5 cm}\hspace{1cm}
\epsfig{figure=ord_orb.eps,width=5.5 cm}\\\vspace{0.7cm}
\epsfig{figure=pss_proc.eps,width=5.5 cm}\hspace{1cm}
\epsfig{figure=sali_2orb.eps,width=5.5 cm}
\end{center}
\vspace*{0.1cm} \caption{Projections of the orbits (i) and (ii) in
the $(x,y)$-plane (panels a), b)). PSS of these two orbits in panel
c). Behaviour of the SALI of the same orbits (panel d)). }
\end{figure*}

\section{Conclusions}
In this paper, we applied the SALI method in the Ferrers barred
galaxy models of 2 and 3 dof. We presented and discussed our results
comparing the index with traditional methods, such as the PSS method
for the 2 dof and showed its effectiveness. We also, calculated
percentages of chaotic and regular orbits and how they change with
the main model parameters, for the 3 dof case.

 \acknowledgements{Thanos Manos was partially supported by
"Karatheodory" graduate student fellowship No  B395 of the
University of Patras and by "Marie –- Curie" fellowship No
HPMT-CT-2001-00338.}

\vfill

\begin{thebibliography}{}{
\bibitem{sk:1}Skokos Ch., 2001, J. Phys. A: Math. Gen., 34, 10029
\bibitem{sk:2}Skokos Ch., Antonopoulos C., Bountis T. and Vrahatis, M., 2004, J. Phys. A, 37, 6269
\bibitem{VKS}Voglis N., Kalapotharakos C. and Stavropoulos I., 2002, Mon. Not. R. Astron. Soc. 337, 619
\bibitem{}Athanassoula E., Bienayme O., Martinet L. and Pfenniger D., 1983, Astr. Astroph. 127, 349
}
\end{thebibliography}
\end{document}